\documentstyle[11pt,paspconf,epsf]{article}

\begin{document}

\title{Binary Evolution and the Progenitor of SN 1987A}
\author{Philipp Podsiadlowski}
\affil{Oxford University, Oxford OX1 3RH, U. K.}

\begin{abstract}
Since the majority of massive stars are members of binary systems, an
understanding of the intricacies of binary interactions is essential for
understanding the large variety of supernova types and subtypes. I 
therefore briefly review the basic elements of binary evolution theory and
discuss how binary interactions  affect the presupernova structure of 
massive stars and the resulting supernovae.\par\medskip

SN~1987A was a highly anomalous supernova, almost certainly because of
a previous binary interaction. The most likely scenario at present is
that the progenitor was a member of a massive close binary that
experienced dynamical mass transfer during its second red-supergiant
phase and merged completely with its companion as a consequence. This
can naturally explain the three main anomalies of SN~1987A: the blue
color of the progenitor, the chemical anomalies and the complex
triple-ring nebula.
 
\end{abstract}

\keywords{SN 1987A, binary interactions, mass transfer}

\def\msun{\hbox{$\,M_\odot $}}
\def\lsun{\hbox{$\,L_\odot $}}
\def\rsun{\hbox{$\,R_\odot $}}

\section{Introduction}

In the last few years, it has become increasingly clear that SN~1987A
in the Large Magellanic Cloud (LMC) was a remarkable, but highly
unusual event. One of the major surprises of SN~1987A was the fact
that the star that exploded was a blue supergiant rather than a red
supergiant as had been predicted. While there have been many attempts
in the early years to explain the blue-supergiant progenitor within
the framework of single stellar evolution theory (e.g., Woosley,
Pinto \& Ensman 1988; for a review see Podsiadlowski 1992), these
models are no longer viable with the best, up-to-date input physics
(Woosley 1997). In particular, the recent large increase in stellar
opacities (Rogers \& Iglesias 1992) had as an immediate consequence
that there are no longer any plausible parameters for which a massive
star first becomes a red supergiant and then experiences a late blue
loop after helium core burning (Woosley 1997). Of course, even if such
models could be constructed, they still would not be promising models
for the progenitor of SN~1987A, since they still could not explain any
of the other major anomalies of this event: the complex triple-ring
nebula surrounding the progenitor and the various chemical anomalies
(see Podsiadlowski 1992 and section~3). It is now almost certain
that these anomalies are somehow connected to binary evolution. In
section~2, I therefore first review some of the main, relevant
principles of binary stellar evolution theory and, in section~3,
summarize the observational constraints any model for the progenitor
has to fulfill.  In section~4, I present in detail an updated version
of a merger scenario that at present provides the only framework in
which not only some, but all of these anomalies can be
understood. Throughout this review, I emphasize the still substantial
theoretical uncertainties in this model and indicate how observations
may help to conclusively verify it in all its details.

\section{Binary interactions}

As is not very widely known in the supernova community, most stars in the
sky are actually members of binary (or multiple) systems. To zeroth order,
{\em all} stars are members of binaries (see, e.g., the references in
Podsiadlowski, Joss \& Hsu 1991 [PJH] and Ghez 1996). Of course to have its
presupernova structure altered, a star has to be in a close,
interacting binary where at least one star fills its Roche lobe during
its evolution. The fraction of massive stars in interacting binaries
can be estimated to be in the range of 30$\,$--$\,50\,$\%. For
example, Garmany, Conti \& Massey (1980) found that $36\,$\% of massive stars
are spectroscopic binaries with massive companions with periods less
than $\sim 1\,$yr. This estimate would imply a {\em true}
interacting-binary frequency of around $50\,$\%. It is worth noting
that Roche-lobe overflow occurs more frequently in evolved phases
simply because the radius of a star expands only by a factor of $\sim
2$ during the main sequence, while it expands by a factor of $\sim 100$
subsequently. This means that, for any plausible orbital-period
distribution, a star is much more likely to encounter mass transfer after
its main-sequence phase. Since stars spend the largest fraction of their
lives on the main sequence, most stars observed in the sky have not
(yet) experienced a binary interaction. On the other hand, supernovae
probe the very final stage in the evolution of a star. Therefore a large
fraction of all supernova progenitors are affected by a previous binary
interaction. This is, at least in part, responsible for the 
large variety of observed supernova types.

In general, two qualitatively very different modes of mass transfer
can be distinguished: more-or-less conservative mass transfer and
dynamically unstable mass transfer.

{\em (Quasi-)conservative mass transfer} usually occurs when the
mass donor has a radiative envelope and the mass ratio of the
mass-accreting to the mass-losing component is not too small (e.g.,
de Loore \& de Gr\`eve 1992). Then, a large fraction
($\ga 0.5$) of the mass lost from the primary is accreted by
the secondary. Thus in this case, both components of the binary are
affected, one by losing mass, the other by accreting it. During mainly
conservative mass transfer, the orbital period tends to increase.

{\em Dynamical mass transfer} usually takes place when the secondary
is a giant star with a deep convective envelope. In this case, mass
transfer is dynamically unstable and leads to the formation of a
common envelope surrounding the core of the giant and the secondary
(Paczy\'nski 1976).  Due to friction between this immersed binary with
the common envelope, the orbit of the binary starts to shrink.
Dependent on how much energy is released in the orbital decay of the
binary and is deposited in the envelope, two different outcomes are
possible. If the deposited energy exceeds the binding energy of the
envelope, the common envelope can be ejected, leaving a very close
binary consisting of a helium star (or Wolf-Rayet star) and a normal
companion star (which is hardly affected by the common-envelope
phase).
If the energy is not sufficient to unbind
the envelope, the less dense component of the immersed binary will
ultimately be tidally destroyed and the two components merge completely
to form a single, but rapidly rotating giant (see section~4.2).

In a typical binary scenario, a binary system may 
experience several different phases of
mass transfer, which is the reason for the large variety of possible
binary scenarios. In the following, I will concentrate on the main
consequences of the various mass-transfer types for the structure
of the immediate supernova progenitors (for a more detailed discussion
see PJH and Hsu et al.\ 1996).

\subsection{Mass loss}

Some 30$\,$--$\,50\,$\% of all massive stars experience Roche-lobe
overflow and mass loss at some point during their evolution. 
In most cases, they lose all of their
hydrogen-rich envelopes and become helium stars or Wolf-Rayet stars,
which are excellent candidates for type Ib/Ic supernovae. In some
cases, however, it is possible for the primary to retain part of
its hydrogen-rich envelope, in particular when mass transfer occurs
very late during the evolution of the primary and when the initial mass
ratio is very close to one. However, even in this case, at most a few
solar masses can be retained in the envelope. The immediate supernova
progenitor will be a {\em stripped supergiant} with a small envelope
mass (Joss et al.\ 1988). 
It is interesting to note that the outer appearance of this
star will be very similar to that of a star that has lost no mass at
all, since the radius and the luminosity of the star are almost
independent of the envelope mass (PJH). The
resulting supernova will, however, be quite different. The light curve
will show no extended plateau phase but resemble a type II-L
or, in the most extreme case, a type IIb supernova (Nomoto et al.\ 1993;
Podsiadlowski et al.\ 1993; Woosley et al.\ 1994; Hsu et al.\ 1996). 

\subsection{Mass accretion}

Just as the structure and appearance of the mass-losing star can be
strongly affected, the structure and further evolution 
of the accreting companion can also be dramatically altered by mass transfer.
This depends, however, strongly on the evolutionary stage of the
accreting secondary at the beginning of the mass-transfer phase.
\par\medskip
\noindent {\em Secondary on the main sequence}\par\noindent
If the secondary is still on the main sequence at the beginning of
the mass-transfer phase, the secondary is usually {\em
rejuvenated} and will behave subsequently (after the mass-transfer 
phase) like a single,
but now more massive star (Hellings 1983; PJH).
However, as was shown by Braun \& Langer (1995), this need
not be the case if accretion occurs very late on the main sequence
and if semi-convection is very slow (combined with the Ledoux
criterion for convective instability), since the convective core will
then not grow significantly as a result of accretion. The subsequent
evolution will  resemble the evolution of a star that accreted after the
main-sequence phase (see below); in particular, the star may never 
become a red supergiant and spend the whole post-main-sequence phase
in the blue-supergiant region of the Hertzsprung-Russell (H-R)
diagram.
\par\medskip\noindent
{\em Secondary has completed hydrogen core burning}\par\noindent
If the secondary has already left the main sequence before the
beginning of the mass-transfer phase, the subsequent evolution will
generally differ quite substantially 
from the evolution of a normal single star. 
Since the mass of the helium core will not grow as a result
of mass accretion, the main effect of post-main-sequence accretion is
to increase the envelope mass relative to the core mass. As was first
shown by Podsiadlowski \& Joss (1989) (see also De Loore \&
Vanbeveren 1992; PJH), this has the consequence that
the star will now not become a red supergiant or, if it was a red
supergiant at the time of the accretion phase, leave the
red-supergiant region and spend its remaining lifetime as a blue
supergiant (see also Barkat \& Wheeler 1989 for a related scenario).
The final location of the star in the H-R diagram
at the time of the supernova depends on how much mass has been
accreted in the accretion phase. The star will be the bluer, the more mass
it has accreted. The final supernova will be of the SN~1987A variety 
(type II[blue]). Since a star has to accrete only a few percent of its
own mass from a binary companion to be spun up to critical rotation,
the progenitor is expected to pass through a phase where it is 
rapidly rotating. Rapid rotation may also induce large-scale mixing in the
accreting star (even across chemical gradients). 
However, whether this happens depends critically on the angular-momentum
transport inside and the structure of the accreting star.

\subsection{Binary mergers}

The most dramatic type of binary interaction is dynamical mass
transfer leading to a common-envelope and spiral-in phase. If the
orbital energy released during the spiral-in is sufficient to eject
the common envelope, the end product is a short-period binary
consisting of a Wolf-Rayet primary and a normal stellar companion.
The primary will eventually explode as a type Ib/Ic supernova. If the
system remains bound, the secondary is also likely to experience
mass loss in the future leading to a second mass-transfer phase, just
as in the mass-accretion scenarios discussed in the previous
subsection.

If the common envelope is not ejected, the two components will merge
completely to produce a more massive and rapidly rotating
single star. This end product resembles in many ways the outcome
of the post-main-sequence accretion models in section~2.2. In
particular, it may end its evolution as a blue supergiant rather than
as a red supergiant either because of the added mass in the envelope
(Podsiadlowski, Joss \& Rappaport 1990) or because of the dredge-up
of helium (Hillebrandt \& Meyer 1989) or both.
The resulting supernova may again be of the SN~1987A variety.
Podsiadlowski et al.\ (1991) have estimated that the combined frequency
for a blue-supergiant progenitor in either an accretion or a merger
scenario is about 5$\,$\% of all core-collapse supernovae with an uncertainty
of about a factor of two.
\par\medskip

%{\noindent\it Complications}\par\medskip\noindent

\noindent{\em Complications.} 
In this section, I only discussed the main types of interactions.
However, many binaries experience more than one phase of mass transfer.
For example, low-mass helium stars, produced in a first mass-transfer phase,
will expand again after core helium burning to become helium giants 
and may fill their Roche lobes for a second time (so-called case BB 
mass transfer; De Gr\`eve \& De Loore 1977; Delgado \& Thomas 1981).
This still does not exhaust the whole variety of multiple 
mass-transfer scenarios. Nomoto {et al.}\ (1994) 
summarized various scenarios to produce bare CO cores as progenitors
of type Ic supernovae, which involve up to {\it three} mass-transfer phases.
\par

\section{Observational Constraints on the Progenitor of SN~1987A}

\subsection{The blue-supergiant progenitor}

One of the major early surprises of SN~1987A was the fact that its
known progenitor, \mbox{SK~$-69^{\circ}$202}, was a blue supergiant
rather than a red supergiant, the normally expected supernova
progenitor. Moreover, from the dynamical age of the surrounding
low-velocity nebula ($\sim 30000\,$yr) one can infer that it was a red
supergiant just a few $10^4\,$yr ago. Any model of the progenitor has
to explain this recent transition {\em and} be consistent with the
general behaviour of massive stars in the LMC in the H-R diagram.
Many of the early models already failed this obvious test (see
Podsiadlowski 1992).

\subsection{The triple-ring nebula} 

One of the most spectacular features of SN~1987A is 
the complex, but very axisymmetric nebula surrounding
the supernova, which was formed out of material that was ejected from
the progenitor in the not-too-distant past. The main geometry of the
nebula consists of three rings, an inner ring centered on the
supernova (Wampler et al.\ 1990; Jacobson et al.\ 1991) and two rings
displaced to the South and the North, but in approximate alignment
with the symmetry axis of the inner ring (Wampler et al.\ 1990;
Burrows et al.\ 1995).
This geometry implies an axisymmetric, but highly non-spherical
structure of the envelope of the progenitor and/or its winds.
The origin of this non-sphericity provides a severe constraint
for models of the progenitor. A plausible mechanism to produce
the required asymmetry is the flattening of the progenitor's envelope
caused by rapid rotation (Chevalier \& Soker 1989). However,
straightforward angular-momentum considerations show that a single
star which was rapidly rotating on the main sequence would be a slow
rotator in any subsequent supergiant phase and could not possibly 
be significantly flattened at the time of the supernova explosion.
Recent HST observations (Pun 1997) which show that the supernova ejecta
themselves are elongated along the symmetry axis of the nebula
provide further evidence for a rapidly rotating, flattened (oblate) 
progenitor, since in this case the supernova shock will propagate faster
in the polar directions, thereby generating a prolate structure
of the ejecta.

\subsection{The chemical anomalies}

The third major surprise of the supernova involves a number of
chemical anomalies in the progenitor's hydrogen-rich envelope
and in the presupernova ejecta.\par\medskip

{\noindent\em The inner ring: the helium anomaly}\par\nobreak\medskip\noindent
As is now firmly established, the composition of the inner ring shows
that significant amounts of CNO-processed material have been dredged up to
the surface (with N/C$\,\sim\,$5, N/O$\,\sim\,$1 [all ratios are by number]) 
and that, most importantly, the helium abundance in the inner ring is 
about twice solar (He/H$\,\sim\,$0.25; H\"oflich 1988; Allen, Meikle \& 
Spyromilio 1989; Fransson et al.\ 1989; Wang 1991; Lundqvist \& Fransson 1996;
Sonneborn et al.\ 1997).

\medskip
{\noindent\em The outer rings: two dredge-up phases?}
\par\medskip\noindent
Recently, Panagia et al.\ (1996) found that the chemical composition
of the outer rings is significantly different from the composition of
the inner ring, indicating a smaller amount of dredge-up of
CNO-processed material (N/C$\,\sim\,$2; N/O$\,\sim\,$0.6). This, if confirmed,
might suggest that there were two dredge-up phases, the first
associated with the normal convective dredge-up when a star becomes a
red-supergiant and develops a deep convective envelope and the second,
a few $10^4\,$yr ago, connected with the event that produced all the
other anomalies as well. Determination of the helium abundance of the
outer rings could shed more light on the details of this process.
\par\medskip
{\noindent\em The barium anomaly}
\par\medskip\nobreak\noindent
The third chemical anomaly, most people had hoped would go away, is the
enhancement of barium (by a factor of 5$\,$--$\,$10) and other
s-process elements in the progenitor's envelope
(Williams 1987; H\"oflich 1988; 
Mazzali, Lucy \& Butler 1992; Mazzali \& Chugai 1995). This suggests
the simultaneous occurrence of hydrogen and some helium-burning
reactions (in particular, C$^{13} + \alpha$) in the outer
parts of the core, similar to the process that produces these elements
in S stars on the asymptotic-giant branch (Sanders 1967).

All of these anomalies together suggest that there was a single,
dramatic event a few $10^4\,$yr ago that is responsible for them.
Several of these show clear fingerprints of binary interactions.
Just as in any good mystery story, there are plenty of traces and clues,
and it is just up to us to decipher their meanings and to reconstruct what
{\em really} happened to the progenitor of SN~1987A.

\section{The Progenitor of SN~1987A: a Binary Merger}
\begin{figure}[p]
\plotone{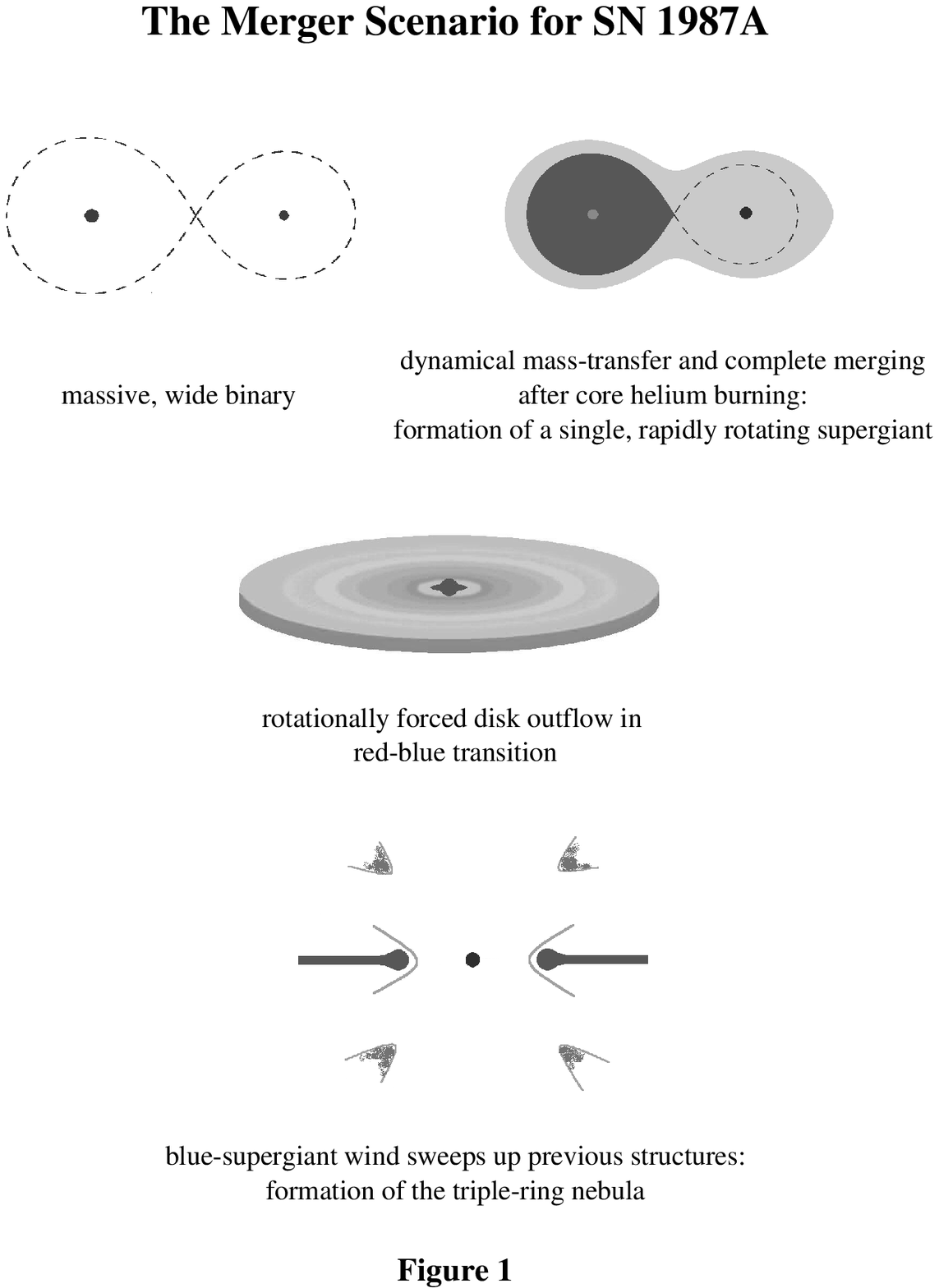}
\end{figure}

Over the last ten years, numerous binary models have been suggested
(for a review, see Podsiadlowski 1992; see also Rathnasree 1993;
Braun \& Langer 1995). Both accretion and merger
models (see section~2) can explain a rapidly rotating blue-supergiant
progenitor.  However, since in the former class of models it is the
secondary that accretes from a more evolved star and since all of this
should have occurred in the very recent past, this would require the
masses of the stars to have been extremely close initially. This
implies enormous fine-tuning of the binary parameters and therefore is
no longer a favoured model.  It now appears most likely that the
progenitor was a binary that merged with its companion in the recent
past (as originally proposed independently by Hillebrandt \& Meyer
[1989] and Podsiadlowski {et al.}\ [1990]; also see Chevalier \& Soker
[1988] for an earlier suggestion).

In this section, I will present an updated version of the merger
scenario outlined in Podsiadlowski (1992) and schematically
illustrated in Figure~1 (details will be published in Podsiadlowski
1997). The calculations use the chemical abundances
for massive LMC stars determined by Russell \& Bessell (1989) and
Russell \& Dopita (1990) (with $Z=0.01$ but an uncertain carbon
abundance), updated opacities by Rogers \& Iglesias (1992) and
Alexander (1994), as provided by Eggleton (1997).

Initially, the progenitor was a member of a very typical binary,
consisting of a primary of $\sim 15\,M_{\odot}$ and a significantly
less massive secondary (5$\,$--$\,10\,M_{\odot}$) in a fairly wide
orbit (with an orbital period of $\sim 10\,$yr), so that the primary
started to fill its Roche lobe only on the asymptotic-giant branch
(i.e., on its second ascent of the red-supergiant branch after helium
core burning). The mass of the companion is not well determined by the
model at the present time. Indeed, a relatively low-mass companion
could be sufficient.  The system then experienced dynamical mass
transfer, leading to the complete merger of the two stars. The end
product of this evolution is a single, but very rapidly rotating red
supergiant, which has been thoroughly stirred up during the merging
process (explaining the main chemical anomalies, provided that this
environment allows for s-processing). The star will now want to shrink
to become a blue supergiant, producing a rotationally forced disk-like
outflow in the process. In the subsequent blue-supergiant phase, the
energetic blue-supergiant wind sweeps up all the structures
generated previously and produces the triple-ring nebula (for a model
of the outer rings in a merger scenario, see Podsiadlowski, Fabian \&
Stevens 1991; Lloyd, O'Brien \& Kahn 1995; Podsiadlowski \& Cumming
1995).\par

\begin{figure}[t]
\includegraphics{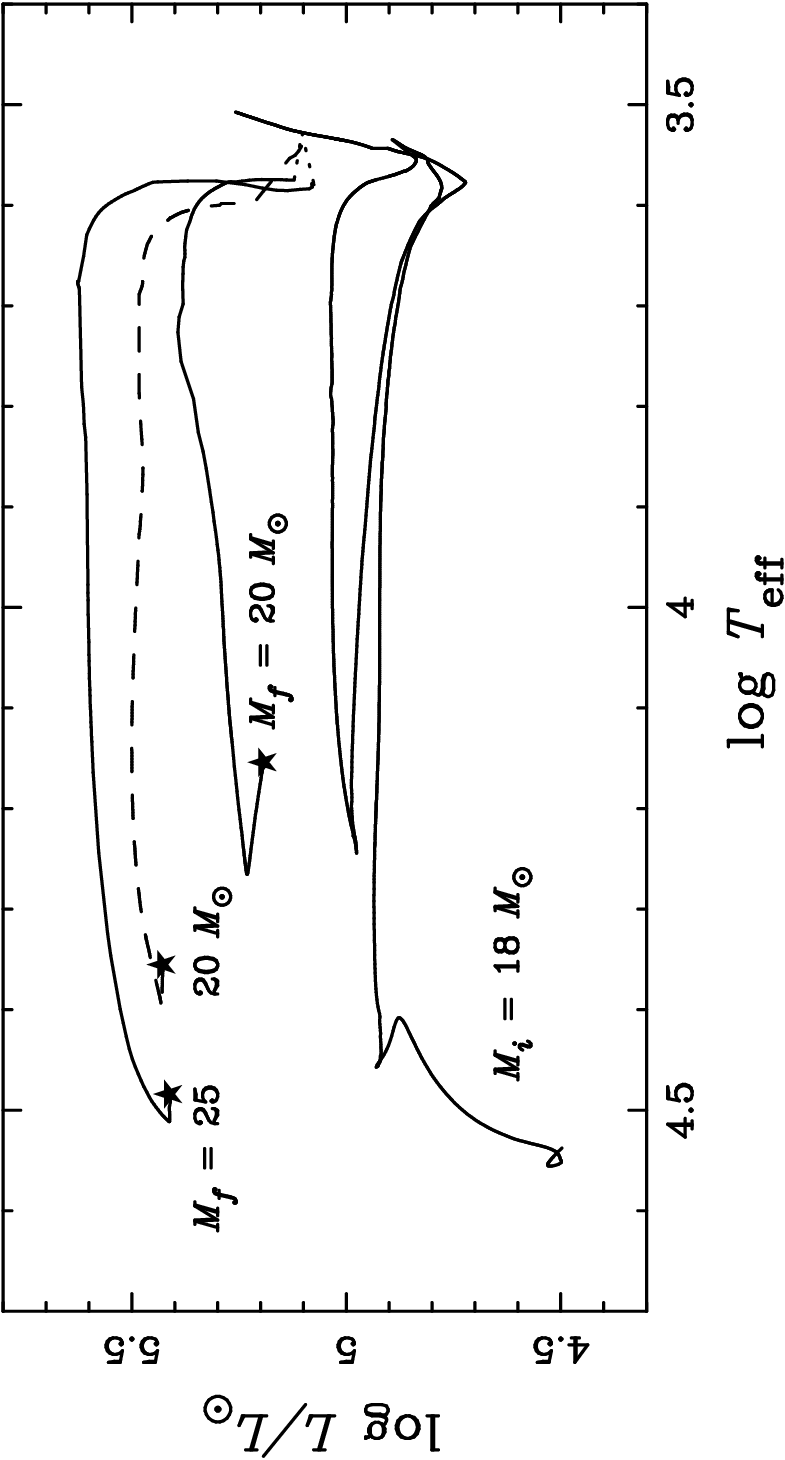}
\vspace{2.6truein}
\noindent{{\bf Figure 2.} 
Representative merger calculations for a star with an initial mass of
18\msun\ merging with a 10\msun\ star after helium core burning for different
amounts of helium dredge-up (solid and dashed curves) and final
masses as indicated.}
\vspace{-0.1in}
\end{figure}

Figure~2 shows several representative merger calculations based on
an analytic theory for the merger process 
(Podsiadlowski 1997; and Podsiadlowski
\& Spruit 1997), for a primary with an initial mass of 18\msun\ merging
with a 10\msun\ companion. In the calculations represented by solid
curves, 1.2\msun\ of the helium core is dredged-up during the merger and
the final masses of the merged stars are 20 and 25\msun, respectively (as
indicated). Both stars have surface abundances of helium and the CNO elements 
that are in excellent agreement with the observational constraints
(section~3.3). In the model shown as a dashed curve, the merger is even
more dramatic and the final object would chemically be classified as a
barium star, though this particular model is somewhat too hot and too
luminous for the progenitor of SN~1987A.

\subsection{Formation of the inner ring}

The production of the inner ring requires a disk-like outflow in the
red-supergiant phase. It has been suggested that gravitational
focusing by a non-interacting companion (similar to what may happen in
some planetary nebulae, e.g., Morris 1981) could provide such a
focusing mechanism. However, the constraints on any distant companion
after the supernova are quite stringent, and its mass could not be
larger than $\sim 2\msun$ (e.g., Plait et al.\ 1995). It is easy to
show, using a Bondi-Hoyle-type wind theory, that the fraction of the
red-supergiant wind that is gravitationally affected by the companion
is of order ($v_{\rm orb}/v_{\rm wind})\,[M_2/(M_1+M_2)]^2$, where
$v_{\rm orb}$ and $v_{\rm wind}$ are the orbital velocity of the
secondary and the wind velocity, respectively, while $M_1$ and $M_2$
are the masses of the primary and the secondary, respectively. For
plausible parameters for the SN~1987A progenitor, this fraction cannot
exceed $\sim 5\,$\%, which is completely insufficient to explain the
large observed asymmetry.

On the other hand in a merger scenario, the problem is not a lack of
rotation but an excess thereof (Podsiadlowski 1992; Chen \& Colgate 1995),
since all the orbital angular momentum in the pre-merger binary will be
deposited in the envelope of the progenitor, spinning it up in the process.
Indeed, for the parameters of the model shown in Figure~2, the total orbital
angular momentum ($\sim 9\times 10^{54}\,$erg s) substantially exceeds 
the maximum angular momentum of a dynamically stable blue supergiant 
($\sim 4\times 10^{54}\,$erg s). This immediately implies that the
merged system will pass through a phase of critical surface rotation in the
final red-blue transition, leading to rotationally enforced, equatorial
mass loss. One can obtain a rough estimate for the minimum amount of mass that
needs to be shed in this process by dividing the excess angular momentum
by the specific orbital angular momentum of the initial binary. This
yields
$$\Delta M \sim {\Delta L\over\sqrt{G M D}}\sim 4\msun\,\left({D\over 10 
{\rm AU}}\right)^{-1/2},$$
where $D$ is the characteristic co-rotation radius  
associated with the mass loss (assumed here to be of order the initial
binary separation). 

This estimate implies that one expects at least
several solar masses to be lost after the merger. How this mass loss took
place in detail is not so clear at the moment. It may involve
a rotationally focused wind, as in the standard model for Be-star disks
by Bjorkman \& Cassinelli 
(1993), or dynamical instabilities in a 
critically rotating object (e.g., Durisen et al.\ 1986;
Livio \& Soker 1988; Taam \& Bodenheimer
1991; Chen \& Colgate 1995). In both cases, one would
expect the formation of a disk-like, equatorial structure with a small
radial velocity (as required to explain the low expansion velocity of 
$10\,$km s$^{-1}$ of the inner ring).

In this context it is worth noting that there is an observed system
that seems to have merged in the very recent past and that shows some
similarities to the SN~1987A progenitor: V Hydrae is a carbon-star
giant rotating near break-up (Barnbaum, Morris \& Kahane 1995).  The
system also shows evidence for a disk-like equatorial and a biconical
polar outflow. Barnbaum et al.\ (1995) even suggested that the
carbon-star characteristics of V Hyd may by a direct consequence of
the dredge-up of carbon {\em caused} by the merging process. While
this system is not entirely comparable to the progenitor of SN~1987A,
one may hope to learn more about this relatively unexplored, but not
at all uncommon ($\sim\,$5\%--$\,10\,$\%) phase of binary evolution from V Hyd
and, of course, SN~1987A.

\subsection{The merger phase}

One of the least studied aspects of the merger scenario involves the details
of the final merging of the two binary components, the secondary and
the helium core of the red supergiant, inside the common envelope.
We (Podsiadlowski \& Spruit 1997) have started to look at this process in
some detail (also see Meyer \& Meyer-Hofmeister 1979; Taam \& Bodenheimer
1989, 1991; Terman, Taam \& Hernquist 1994).

The spiral-in phase ends when the embedded secondary starts to fill
its own Roche lobe (at a separation of $\sim 10\rsun$). It then begins
to transfer mass to the helium core. This mass-transfer process
resembles in many respects the ``normal'' mass transfer in interacting
binaries, except that it occurs inside a low-density, opaque
envelope. The mass-transfer rate is determined by the frictional drag
the orbiting binary experiences with the common envelope.  The
timescale for the final destruction of the secondary is uncertain.
Rough guesses range from a few days (assuming that the envelope does
not expand as a result of the spiral-in) to hundreds of years
(assuming that the frictional luminosity is self-regulated at the
Eddington limit; e.g., Meyer \& Meyer-Hofmeister [1979]). In the
calculations presented earlier, I adopted, somewhat arbitrarily, a
timescale of $1\,$yr.  Because of the relatively large size of the
helium core, the stream emanating from the Roche-lobe filling
secondary does not self-intersect and form an accretion disk but
rather impacts with the core and penetrates it. In other words, it
drills a hole into the core and starts to erode it, causing
the dredge-up of helium-rich material. The
penetration depth can be estimated from the condition that the ram
pressure in the stream must be of the order of the ambient pressure in
the helium core. One major uncertainty in this estimate is how much
energy is dissipated inside the stream by internal shocks (which must
occur because of the pressure focusing of the stream and
angular-momentum constraints). The characteristic temperature of the
shock-heated material in the stream-impact region is $T\sim 2\times
10^8\,$K. Since this material is hydrogen-rich, but the temperature
more typical of helium burning, one can expect some unusual
nucleosynthesis in this region, possibly responsible for the more
exotic chemical anomalies of SN~1987A like the barium anomaly.

Preliminary calculations using extended nuclear-reaction networks
show that significant s-processing (with neutron exposures
of $\sim 10^{27}\,$cm$^{-2}$ with C$^{13}+\alpha$ as neutron source) 
is possible in this environment. This requires
that the dredge-up region of the core is continually mixed with
some of the hydrogen-rich envelope that serves as a proton reservoir
for an extended period of time (of order the merger timescale). 
Unfortunately, some of the key reaction rates are very temperature sensitive,
while the burning conditions are not sufficiently well determined by
the present analytic model to allow very firm conclusions.
In addition, these calculations suggest that the abundances of the
rare nitrogen and oxygen isotopes N$^{15}$ and O$^{17}$ may
be overabundant by a large factor. This could provide a potentially
sensitive tracer for the burning conditions in this environment.

\section{Concluding Remarks}

As I have shown in this review, binary interactions are almost certainly
responsible for some of the large variety of observed supernova types
in general and therefore cannot be ignored. As far as SN~1987A is concerned,
a merger scenario provides at present the only model for the progenitor 
that can explain all the major features of this remarkable event. 
There are still many theoretical uncertainties,
in particular involving the details of the merging process and the
associated nucleosynthesis. Observations of abundances in different
parts of the nebula may be particularly useful in helping to reconstruct the
detailed merger history (for example, the helium abundance in the
outer rings). In addition, there may be other chemical anomalies 
(overabundances of N$^{15}$ and O$^{17}$) that may be used as direct
tracers of the unusual burning conditions that occur during the final
merging process. 

While a lot has been learned about the progenitor of SN~1987A in the last
ten years, there still remains much more to be done, both theoretically and
observationally. Ultimately, when the supernova blast wave reaches the
inner ring, at least the structure of the whole nebula will be finally
revealed and confirm or refute some of the aspects of the merger scenario.

\end{document}